\documentclass[aps,showpacs]{revtex4}

\usepackage{amsmath,amsthm}

\usepackage{graphicx}

\newtheorem{lem}{Lemma}
\newtheorem{thm}{Theorem}

\begin{document}

\title{Solving the Richardson equations close to the critical
points}


\author{F. Dom\'{\i}nguez$^{1}$, C. Esebbag$^{1}$ and J. Dukelsky$^{2}$}

\address{$^1$ Departamento de Matem\'aticas, Universidad de Alcal\'a, 28871
Alcal\'a de Henares, Spain. \\$^2$ Instituto de Estructura de la
Materia. CSIC. Serrano 123. 28006 Madrid. Spain.}


\begin{abstract}
We study the Richardson equations close to the critical values of
the paring strength $g_c$ where the occurrence of divergencies
preclude numerical solutions. We derive a set of equations for
determining the critical $g$ values and the non-collapsing pair
energies. Studying the behavior of the solutions close to the
critical points, we develop a procedure to solve numerically the
Richardson equations for arbitrary coupling strength.
 \end{abstract}


 \maketitle

\section{Introduction}
\label{intro}

The exact solution of the BCS or pairing Hamiltonian was presented
by Richardson in series of papers beginning in 1963 \cite{Rich0,
Rich1}, just a few years after the seminal paper of Bardeen,
Cooper and Schrieffer \cite{BCS}. Despite of the new avenues of
research that it could have opened at this early stage in the
development of the theory of superconductivity, the work passed
almost unnoticed due, perhaps, to the technical difficulties in
solving numerically the set of non linear equations for the
spectral parameters. In recent year, the Richardson exact solution
was rediscovered and applied successfully to ultrasmall metallic
grains where it was shown to be essential for the description the
soft crossover between superconductivity and the paring
fluctuation regime as a function of the grain size \cite{grains}.
Since then the Richardson model has been extended to a wide class
exactly solvable models called Richardson-Gaudin (RG) models with
potential applications to several quantum many body fermion and
boson systems \cite{duke1, duke2, Links, duke3}. However, the
numerical treatment of the exact solution for moderate to large
size systems is still a cumbersome task in spite of the recent
efforts to overcome this problem.

For the sake of clarity, let us begin by introducing the
Richardson equations for an $M$ fermion pair system:

\begin{equation}
1- 4g\sum_{j=1}^{L}\frac{d_{j}}{2\eta _{j}-e_{\alpha }}+
4g\sum\limits_{\scriptstyle \beta  = 1  \atop
  \scriptstyle (\beta \ne \alpha)}^M\frac{1}{e_{\alpha
}-e_{\beta }}=0, \label{richA}
\end{equation}
where $g$ is the pairing strength, $\eta_j$ are a set of $L$
parameters usually related to the single particle energies, $d_j$
are the effective degeneracies defined as $d_j = \nu_j/2 -
\Omega_j/4$ with $\nu_j$ de number of unpaired fermions in level
$j$ (Seniority quantum number) and $\Omega_j$ the total degeneracy
of the level $j$. The $e_\alpha$'s are the $M$ unknowns parameters
called pair energies. Given the $L$ parameters $\eta_j$, the $L$
effective degeneracies $d_j$ and a pairing strength $g$, the pair
energies $e_\alpha$ are obtained by solving the set of $M$
nonlinear equations (\ref{richA}). However, it is in general not
easy to find a good initial guess that would lead directly to the
appropriate solution. Instead, it is customary to begin with a
given
 configuration in the weak coupling limit ($g\rightarrow 0$) where
the equations (\ref{richA}) can only be satisfied for $e_\alpha
\rightarrow 2\eta_j$. The exact solution is then evolved step by
step for increasing values of $g$ up to the desired value
\cite{RichA}.
 In each step one uses the
solution obtained in the previous step as the input data for the
numerical solver. Even if the procedure is successful, leading to
the correct solution, it involves a heavy numerical work. In most
cases, this procedures is stopped due to the existence of
singularities for some critical values $g_c$ of the pairing
strength. At $g=g_c$ a subset of pair energies $e_{\alpha }$ turn
out to be equal to $2\eta _{k}$ for some $k$, giving rise to
divergencies in some terms of eq. (\ref{richA}) \cite{RichB}.
Remarkably, these divergencies cancel out and the corresponding
solution (the set of $e_{\alpha }$) is a continuous function of
$g$ in a neighborhood of $g_c$. In fact, the solutions are always
continuous for every value of $g$.

The existence of such critical points has significant consequences
in the numerical solution of the equations which becomes unstable
near $g_c$. Very slow convergency, no convergency at all or
jumping to another solution are the typical problems found close
to $g_c$. Moreover, there may be more than just one critical value
in the real interval  $[0,g]$, and we do not know \emph{a priori}
where the $g_c$'s are located. Those values are known only after
one has already solved the equations for a large number of points
surrounding the $g_c$'s and making some kind of interpolation
afterward. Therefore, the localization of the critical values of
$g$ relies on heuristic procedures and is not based on any
mathematical properties of the equations.

An alternative procedure to cross the critical region, based on a
non linear transformation of the collapsing pair energies was
recently proposed \cite{Romb}. However, this procedure is unable
to predict the critical values of $g$.



 In this paper we will analyze the properties of Richardson
 equations in the vicinity of critical $g$ values. We will derive
conditions that allow to \emph{a priori} determine all the
critical values $g_c$ associated to any ``single particle level''
$\eta_j$, and provide the exact solution at these points. In
addition, we will describe the asymptotic behavior of the solution
in  the limit $g\rightarrow g_c$, and we will present an algorithm
to solve the Richardson equations for values of $g$ near any
$g_c$.

\section{Transforming the Richardson Equations. The Cluster Equations}

Our approach to deal with the singularities in eq. (\ref{richA})
consists in transforming the system through a change of variables
first suggested by Richardson \cite{RichB} and later used by
Rombouts {\it et al.} \cite{Romb} to develop a numerical algorithm
for solving the equations near the critical points. We have
already mentioned  that for  critical values of $g$ some subset of
the pair energies becomes equal to one of the values of $2
\eta_k$. It can be shown \cite{RichB} that the number of such pair
energies is
 $M_k=1-2 d_k$. We can then characterize a critical
point $g_c$ by the condition $\lim_{g \rightarrow g_c} e_{\alpha
}= 2 \eta_k$ $\forall \alpha \in C_k$, where $C_k$ stands for the
subset of indices of the $M_k$ pair energies that satisfy that
limit, i.e. the $e_\alpha$'s that cluster around the real point $2
\eta_k$ in the complex plane for $g$ near $g_c$ (see
\cite{electro} for a graphical representation of these clusters).
 Therefore, we will deal with two subsets of variables, the $M_k$
$e_\alpha$'s with $\alpha \in C_k$ that give rise to the
singularities, and the remaining $(M-M_k)$ variables with $\alpha
\notin C_k$ which behave smoothly close to $g_c$. Consequently, we
will treat separately the $M_k$ Richardson equations with $\alpha
\in C_k$ (the cluster equations):

\begin{equation}
1-4g \frac{d_{k}}{2\eta _{k}-e_{\alpha }}
-4g\sum\limits_{\scriptstyle j = 1  \atop
  \scriptstyle (j \ne k) }^{L}\frac{d_{j}}{2\eta _{j}-e_{\alpha }}+
4g\sum\limits_{\scriptstyle \beta \in C_k  \atop
  \scriptstyle (\beta \ne \alpha)}\frac{1}{e_{\alpha
}-e_{\beta }} +4g\sum\limits_{\beta  \notin C_k}\frac{1}{e_{\alpha
}-e_{\beta }}=0. \label{eqcluster}
\end{equation}

The second and fourth terms of eq. (\ref{eqcluster}) diverge for
$g \rightarrow g_c$ since $(2\eta _{k}-e_{\alpha })$ and
$(e_{\alpha }-e_{\beta })$ go to zero. Moreover these quantities
must approach zero at the same rate in order to cancel out. To avoid the
singularities we will multiply the cluster
equations by $(2\eta _{k}-e_{\alpha })^p$ (for some $p>0$). With
this in mind we will introduce a change of variables for the pair
energies in the cluster (see ref. \cite{RichB, Romb})
\begin{equation} \label{sp}
S_p=\sum_{\alpha \in C_k}(2\eta _{k}-e_{\alpha })^p \qquad
\mbox{for} \qquad p=1,2,\ldots,M_k .
\end{equation}

This is an invertible transformation and, in principle, we can
always recover the $e_\alpha$'s for any arbitrary set of $S_p$.
Keeping in mind that we have just $M_k$ independent variables
$S_p$ we will extend the definition allowing $p$ to be any
positive integer or zero (note that $S_0=M_k$). The variables
$S_p$ behave smoothly in the vicinity of $g_c$, they are real and,
for $p>0$, they satisfy  $\lim_{g\rightarrow g_c}S_p=0$.

In order to transform the cluster equations (\ref{eqcluster}) we
first multiply them by $(2\eta _{k}-e_{\alpha })^p$, for a generic
integer power $p$, and then sum the resulting equations over
$\alpha \in C_k$:
\begin{equation}
S_{p}+ \, 2g  p \, S_{p-1}-2g \sum_{i=0}^{p-1}{S_{p-i-1}S_{i}} \\
-4g\sum\limits_{\scriptstyle j = 1 \atop
  \scriptstyle (j \ne k) }^{L}\sum_{\alpha \in C_k}
  \frac{d_{j}(2\eta _{k}-e_{\alpha })^p}{2\eta _{j}-e_{\alpha }}
  +4g\sum\limits_{\beta  \notin C_k}\sum_{\alpha \in C_k}
  \frac{(2\eta _{k}-e_{\alpha })^p}{e_{\alpha}-e_{\beta }}=0 .
  \label{eqcluster2}
\end{equation}

 Note that the last two terms in the left hand side of the
equation cannot be easily rewritten in terms of $S_p$ if we
restrict to values of $p \le M_k$. To overcome this limitation, we
will make a series expansion of those terms, allowing $p$ to be
any integer. We are interested in the limiting behavior of the
solution close to $g_c$ so that our reasoning will be valid in an
interval $g\in (g_c-\delta,g_c+\delta)$ for some small enough
radius $\delta>0$. Taking into account that $(2\eta _{k}-e_{\alpha
})$ is infinitesimal at $g_c$ and that $e_{\beta }$ in equation
(\ref{eqcluster2}) does not belong to the cluster, we have:

\begin{eqnarray}
\frac{1}{2\eta _{j}-e_{\alpha }} &=& \frac{1}{2\eta _{j}-2\eta
_{k}}\sum_{n=0}^{\infty}\left(\frac{2\eta _{k}-e_{\alpha }}{2\eta
_{k}-2\eta _{j}}\right)^n ,  \nonumber \\
\frac{1}{e_{\alpha}-e_{\beta }}  &=& \frac{1}{2\eta_{k}-e_{\beta
}} \sum_{n=0}^{\infty} \left(\frac{2\eta_{k}-e_{\alpha }}{2\eta
_{k}-e_{\beta }}\right)^n .\label{series}
\end{eqnarray}

These are absolutely (and rapidly) convergent geometric series, a
property that will remain valid after a finite summation over
$\alpha$. Introducing eq. (\ref{series}) in eq. (\ref{eqcluster2})
and making the summation over $\alpha$ we get for $p=1$:

\begin{equation}
S_1-4g\, d_k \, M_k-2g(M_k^2-M_k)+4g \sum_{n=0}^{\infty}
S_{1+n}P_n=0 ,\label{eqs1}
\end{equation}
where the $\{P_n\}$ is a set of coefficients that will be defined
below. Taking in (\ref{eqs1}) the limit $g\rightarrow g_c$ the
second and third terms must cancel out and one gets the cluster
condition $M_k=1-2 d_k$. Similarly, for $p>1$ we obtain:
\begin{equation}
S_p-2g(M_k+1-p)S_{p-1}-2g
\sum_{i=1}^{p-2}S_{p-i-1}S_i+4g\sum_{n=0}^{\infty}S_{p+n} P_n=0 \\
\mbox{ \ for all \ } p>1 \label{eqSp}.
\end{equation}

The coefficients $P_n$ appearing in (\ref{eqs1}) and (\ref{eqSp})
are defined as:
\begin{equation} \label{pn}
P_n=\sum_{\scriptstyle j = 1 \atop \scriptstyle (j \ne k) }^{L}
\frac{d_j}{\left(2\eta _{k}-2\eta_{j}\right)^{n+1}}\, + \,
\sum\limits_{\beta  \notin C_k} \frac{1}{\left(2\eta_{k}-e_{\beta
}\right)^{n+1}} \hspace{.3cm}.
\end{equation}

The infinite recurrence relation (\ref{eqSp}) is the fundamental
equation of our formalism. In order to achieve our main goals we
need to establish some important properties (lemma \ref{lem1} and
lemma \ref{lem2}) of the variables $S_p$ and the series appearing
in eqs. (\ref{eqs1}) and (\ref{eqSp}). It is easy to show that
these series are bounded by their first element $S_p$ through the
condition:

\begin{equation}
\left|\sum_{n=0}^{\infty}S_{p+n} P_n\right|\le\left|S_p\right| \,
D \, \frac{1}{1-x} ,
\end{equation}
where $D$ is a positive bounded number involving  $\eta_j$ and
$e_{\beta}$, and $x$ is an upper bound for the absolute values of
the geometric series ratios in eq.(\ref{series}) satisfying
$\lim_{g\rightarrow g_c}x=0$. On the other hand, we have that
$|S_p| \le \sum_{\alpha \in C_k} |2\eta_k-e_{\alpha}|^p$ with
$(2\eta_k-e_{\alpha}) \rightarrow 0$ as $g \rightarrow g_c$.
Therefore, we will assume in what follows that there is an integer
number $n_{\delta}$ big enough such that we may disregard every
term $S_p$ (and related series) for $p>n_{\delta}$.

The next aspect we need to analyze is the order of the
infinitesimal $S_p$ in relation to the order of $S_1$. We will
demonstrate that $S_1,S_2, S_3, \ldots,S_{M_k}$ are of the same
order, implying that $\lim_{g\rightarrow g_c} S_p/S_1=\chi_p \ne
0$ for $p \le M_k$. On the other hand, $S_p$ is of higher order
than $S_1$ if $p> M_k$, that is $\lim_{g\rightarrow g_c}
S_p/S_1=0$. In lemmas \ref{lem1} and \ref{lem2} we will
demonstrate both assertions.

\begin{lem} \label{lem1}
Assuming that $(1+4 g P_0)\ne 0$, with $g\in
(g_c-\delta,g_c+\delta)$ and $P_0$ defined in (\ref{pn}),  the set
$ \left\{S_2,S_3,\ldots,S_{\mu} \right\}$, for some positive
integer $  \mu $ with $ 2 \le \mu \le M_k $, are infinitesimal of
the same order as $S_1$ at $g=g_c$. In the general case $ \mu =
M_k $.
\end{lem}

\begin{proof}
Rearranging equation (\ref{eqSp}) we get:
\begin{equation}\label{eqSp2}
S_p=\frac{1}{(1+4 g P_0)} \left[2g(M_k+1-p)S_{p-1}+2g
\sum_{i=1}^{p-2}S_{p-i-1}S_i-4g\sum_{n=1}^{\infty}S_{p+n} P_n
\right].
\end{equation}
To prove our statement we will apply this equation recursively
starting with $p=2$. In this case we get:
\[ 
S_2=\frac{ 2g(M_k-1)}{(1+4 g P_0)}S_{1}-\frac{ 4g}{(1+4 g
P_0)}\sum_{n=1}^{\infty}S_{2+n} P_n ,
\]  
so  that $S_2$ is of the same order as $S_1$. In the next step we
set $p=3$ in eq. (\ref{eqSp2}), getting $S_3$ in terms of $S_2$
and $S_p$'s with $p>3$. Replacing $S_2$ with the value obtained
previously and rearranging the resulting expression we get for
$S_3$:
\begin{eqnarray*}
S_3=\frac{ 1}  {(1+4 g P_0)^2 + 8 g^2 P_1(M_k-2)  }
\left[(2g)^2(M_k-2)(M_k-1)S_{1}+2g(1+4 g P_0)S_{1}^2\right. \\
\left.-4g \sum_{n=1}^{\infty} (2g(M_k-2)P_{n+1}+(1+4 g
P_0)P_n)S_{3+n} \right],
\end{eqnarray*}
from where we see that $S_3$ is of the same order as $S_1$. We can
continue this process replacing in each step $S_{p-1}$ in eq.
(\ref{eqSp2}) for the  expression computed in the previous step
and solving again for $S_p$, getting for every $p>2$ and $p\le
M_k$:
\begin{equation} S_p  = B_p \,
S_1 + \sum\limits_{n = 1}^\infty  {D_n^{(p)}\, S_{p + n} }  +
\sum\limits_{\scriptstyle i,j \hfill \atop
  \scriptstyle i + j < p \hfill} {X_{i,j}^{(p)} }\, S_i\, S_j
\end{equation}
where the coefficients $B_p$, $D_n^{(p)}$ and $X_{i,j}^{(p)}$  can
be calculated recursively.  For intance, for the linear terms we
have the recurrence relation
\begin{eqnarray}
 B_{p + 1}  &=& \frac{{2g(M_k  - p)}}{{(1 + 4gP_0 ) - 2g(M_k  - p)D_1^{(p)} }}B_p ,  \nonumber  \\
 D_n^{(p + 1)}  &=& \frac{{2g}}{{(1 + 4gP_0 ) - 2g(M_k  - p)D_1^{(p)} }}
 \left( {(M_k  - p)D_{n + 1}^{(p)}  - 2P_n } \right) ,
 \label{recrel}
 \end{eqnarray}
starting with the known coefficiens for $p=2$.

In this way we have shown that the local expansion of $S_p$ has
linear terms in $S_1$ and in $S_{p + n}$ plus quadratic terms,
thus we conclude that $S_p$ is an infinitesimal of the same order
as $S_1$ for $1 \le p \le M_k$.

Let us now consider what happens when $p$ takes the value $M_k+1$.
In such a case the factor $(M_k +1 - p)$ in eq. (\ref{eqSp2})
becomes zero and the chain of replacement is broken. The expansion
of $S_{M_k+1}$ has only linear terms of higher order and quadratic
terms involving $S_{1}$. Therefore for $p> M_k$ there are no
linear terms in $S_{1}$. We could have reached the same conclusion
analyzing the recurrence relation (\ref{recrel}).

Finally and for the sake of completeness, let us mention that
after replacing $S_{p-1}$ in equation (\ref{eqSp2}) the terms
involving $S_{p}$ on both sides of the equation might cancel out.
It would be equivalent to the cancellation of the denominators in
the recurrence relation (\ref{recrel}). In such a case our process
of replacement is interrupted and ends for $p=\mu<M_k$.
\end{proof}

So far, we have shown that the first $S_{p}$'s are of the same
order as $S_{1}$. At the same time, is was suggested that $S_{p}$
is of higher order if $p>M_k$. In the next lemma we will proof
this statement.

\begin{lem} \label{lem2}
The set of variables $ \left\{S_p\right\}$ for  $ p> M_k $ are
infinitesimal of
 higher order than $S_1$ at $g=g_c$.
\end{lem}
\begin{proof}
In this case we proceed in the reverse way as we have done in
lemma \ref{lem1}. We will begin with an $S_{p}$ for a  large value
of $p$ and carry out the replacement in backward direction. Once
again we start rewriting equation (\ref{eqSp}) to obtain:
\begin{equation}\label{eqSp3}
S_p=\frac{1}{(M_k-p)} \left[ \frac{(1+4 g P_0)}{2g}
S_{p+1}+2\sum_{n=1}^{\infty}S_{p+1+n} P_n
-\sum_{i=1}^{p-1}S_{p-i}S_i \right].
\end{equation}

We will now take, for  $g\in (g_c-\delta,g_c+\delta)$, an integer
number $n_{\delta}$ large enough such that we may disregard every
term $S_p$  for $p>n_{\delta}$. We start making $p=n_{\delta}$ in
eq. (\ref{eqSp3}). After removing the negligible terms it leads us
to:
\[
S_{n_{\delta}}=\frac{1}{(M_k-n_{\delta})}\sum_{i=1}^{n_{\delta}-1}S_{n_{\delta}-i}S_i
.
\]

In the second step we make $p=n_{\delta}-1$ in eq. (\ref{eqSp3})
and replace the value of $S_{n_{\delta}}$ computed before.
Repeating the same operation for $p=n_{\delta}-2,\,
n_{\delta}-3,\ldots,M_k+1$ and replacing, in each case,  the
linear term for its value computed in the previous step,
eventually we get the following quadratic expression for every
$p>M_k$:
\begin{equation} \label{cuad}
S_{p}=\sum_{i,j  \atop {i+j\ge p} } Y_{i,j}^{(p)}   S_{i}S_j .
\end{equation}

In this equation  the coefficients $Y_{i,j}^{(p)}$ can be
calculated recursively and  the indices $i,\, j$ take values
starting from $1$ but keeping the condition $i+j\ge p$. As only
quadratic terms are retained we conclude that $S_p$ for $p>M_k$
are infinitesimal of higher order than $S_1, S_2,\ldots,S_{M_k}$.
Note once again that the chain of replacement cannot be continued
for $p \le M_k$.\end{proof}

So far we have shown  some essential properties of the variables
$S_p$ and the related series that are required to develop our
formalism. In the next section we will derive the equations that
allow to calculate the values of $g_c$ associated with any single
particle level $\eta_k$.

\section{Critical values of $g$ and
the solution of Richardson Equations}\label{sectiongc}

In this section we will derive a set of equations suitable to
compute all the values of  $g_c$ for any single particle level
$\eta_k$. Furthermore, we will obtain,  at the same time, the
solution of Richardson equation at $g_c$, namely, the values of
the unknowns $e_{\beta}$ for $\beta \notin C_k$, assuming that the
variables in the cluster reach their limiting value $e_{\alpha}=
2\, \eta_k$ \ ($e_{\alpha} \in C_k$). We summarize our results in
the following theorem:

\begin{thm}
 All critical $g$ values of the Richardson Equations (\ref{richA}) associated with the single particle
level $\eta_k$, and the corresponding values of the non collapsing
pair energies $e_{\beta}$ ($\beta \notin C_k$) are the solutions
of the following system of equations:
\begin{subequations}\label{detgc}
\begin{equation}
\left|
\begin{array}{*{8}l}
 (1+4g P_0) &   4g P_1  &  4g P_2 &  \cdots & 4g P_{M_k-1}   \\
   -2g(M_k-1)  & (1+4g P_0) & 4g P_1 &  \cdots& 4g P_{M_k-2}   \\
  0  & -2g(M_k-2) & (1+4g P_0)  &  \cdots& 4g P_{M_k-3} \\
\cdots &   \cdots &  \cdots &\cdots & \cdots\\
0& 0& \cdots & -2g &  \; \; (1+4g P_0)   \\
\end{array} \right|=0
\end{equation}
\begin{equation}\label{eqnocluster}
1-4g\sum_{j = 1}^{L}\frac{d_{j}}{2\eta _{j}-e_{\alpha }}+
4g M_k \frac{1}{e_{\alpha}- 2\eta _{k}} +4g\sum\limits_{\scriptstyle \beta \notin C_k  \atop
  \scriptstyle (\beta \ne \alpha)}\frac{1}{e_{\alpha
}-e_{\beta }}=0 \mbox{\ \ \ for all \ } \alpha \notin C_k
\end{equation}
\end{subequations}

\end{thm}

\begin{proof}

Let us consider the transformed Richardson equations according to
the change of variables defined in eq. (\ref{sp}). This system is
formed by the $(M-M_k)$ Richardson equations (\ref{richA}) for
$\alpha \notin C_k$ together with eq. (\ref{eqs1}) and the subset
of equations (\ref{eqSp}) for $p=2,3,\ldots,M_k$. It is a set of
$M$ independent equations equivalent to the original ones. In
fact, we can consider the unknowns $S_1, S_2,\ldots,S_{M_k}$ as
independent variables together with $\{e_{\beta}\}$ for $\beta
\notin C_k$. The variables $\{e_{\alpha}\}$ for $\alpha \in C_k$,
that still appear in the equations, and the $S_p$ for $p>M_k$ are
formally functions of $S_1, S_2,\ldots,S_{M_k}$ computable through
the inverse of transformation (\ref{sp}).
 At this point we are ready to  analyze  the $M_k$ cluster equations (\ref{eqs1} and
\ref{eqSp} for $p \le M_k$) in the limit $g \rightarrow g_c$,
keeping the lower order terms. According to lemmas (\ref{lem1})
and (\ref{lem2})   we can discard terms with $p>M_k$ and terms
involving the products $ S_i\, S_j$, since they are of higher
order than  $S_1, S_2,\ldots,S_{M_k}$, thus we get the system of
equations:
\begin{equation}\label{linearsys}
\left\{
\begin{array}{*{10}r}
 (1+4g P_0) S_1 \; \; \; \;  + \; 4g P_1 S_2  + \; 4g P_2 S_3  +\; \cdots + \; 4g P_{M_k-1} S_{M_k}  &=&0\\
   -2g(M_k-1) S_1 + (1+4g P_0) S_2  + \;4g P_1 S_3  +\; \cdots+\; 4g P_{M_k-2} S_{M_k} & =&0\\
     \; -2g(M_k-2) S_2  + \;(1+4g P_0) S_3  +\; \cdots +\; 4g P_{M_k-3} S_{M_k}  &=&0\\
 \cdots \hspace{4cm}  &  \cdots &\\
-2g S_{M_k-1} + (1+ 4g P_0) S_{M_k}  &=&0 \\
\end{array} \right.
\end{equation}

This is a linear homogeneous system with the trivial solution
$S_1=0,\; S_2=0,\ldots,\; S_{M_k}=0$, and indeed the variables
$S_p$ vanish for $g=g_c$. Since we are looking for nontrivial
solutions that are continuous functions of $g$ valid also in the
vicinity of $g_c$ where $S_p \neq 0$, the condition for such solutions to exist is
the cancellation of the determinant of the linear system (shown in
eq. \ref{detgc}a). As the coefficients of the system are functions
of the variables $e_{\alpha}$ not belonging to the cluster we have
to resort to the ($M-M_k$) remaining Richardson equations
(\ref{richA}) with $\alpha \notin C_k$. After realizing the limit
$g \rightarrow g_c$, we obtain the system of equations
(\ref{detgc})  introduced by the theorem.
\end{proof}

As we have already stated, the numerical solutions of the
non-linear system of equations (\ref{detgc}) provide all the
critical values $g_c$ for the single particle level $\eta_k$ that
has been previously selected. In addition, for each solution, we
get the complete set of pairs energies $e_{\alpha}$ defining the
wave function of the quantum many body system.

With respect to the possible complex $g_c$ solutions, they may be
still interpreted as critical values for not hermitian integrable
hamiltonians, but we will not analyze these cases in the present
paper.

 Once we have the numerical solution of Eqs. (\ref{detgc}) we can use the values
of $g_c$ and $e_\alpha$'s ($\alpha \notin C_k$) to compute the
coefficients of the homogenous  linear system (\ref{linearsys}).
Solving this linear system yields  (after normalizing to $S_1=1$)
the limit ratio  $\chi_p=\lim_{g \rightarrow g_c}
\frac{S_p}{S_1}$. As $S_p=0$ at $g=g_c$ for all $p \ge 1$,  we can
obtain from eqs. (\ref{linearsys}) the derivatives:
\begin{equation}
 \chi_p=\left. \frac{d \! S_p}{d \! S_1}
\right|_{g=g_c}
\end{equation}


\section{Solving Richardson equations near $g_c$} \label{sectionder}

 So far we have shown how to compute the
solution at $g=g_c$.
We will now derive a method to approach the solution at a value
$g_0$ close to the critical point $g_c$. The main issue in the numerical solution  of the Richardson equations is to determine  a good initial guess for the pair energies, specially in the vicinity of $g_c$ where the equations become unstable. With that initial guess the solution
 at $g=g_0$ is obtained with standard numerical technics. Once we
 have the solution for this specific value $g_0$, we can reach a more
 distant value of $g$, by increasing (or decreasing) $g$, step by
 step, using the solution of the previous step as the starting
 guess for the next one.

 An appropriate set of starting values at  $g_0=g_c+\delta g$ can be obtained by means of  a linear
approximation for the
 parameters  $S_p$ and $e_{\alpha}$:
\begin{align}\label{linaprox}
S_1(g_0) & \approx  \left(\frac{d\!S_1}{dg}\right)_{g=g_c}
\delta\! g , \qquad \quad
  S_p(g_0)  \approx  \left( \frac{d\!S_1}{dg}\right)_{g=g_c} \chi_p \: \delta\! g
\quad (p>1) ,\nonumber
\\[5mm]
e_{\alpha}(g_0) & \approx   e_{\alpha}(g_c) + \left(
\frac{de_{\alpha}}{dg}\right)_{g=g_c} \delta\! g \qquad (\alpha
\notin C_k ),
\end{align}
this approximate values  will
be the initial guess to solve the Richardson equations.

In order to determine the derivative $dS_1/dg$ at $g=g_c$ we will
consider up to second order terms in equations (\ref{eqs1}) and
(\ref{eqSp}) and substitute the variables $S_p$ in terms of the
quadratic expression $S_p \approx \chi_p S_1 + a_p S_1^2$, where
$\chi_p=0$ for $p> M_k$, $\chi_{1}=1$ and $a_1=0$. From equation
(\ref{cuad}) we can see that $S_p$ has quadratic terms in $S_1$
only for $p\le 2M_k$, while for greater values of $p$ the order in
$S_1$ is higher due to the condition $i+j \ge p$. Therefore, we
will retain just the first $2 M_k$ variables $S_p$ and deal with a
system of $2 M_k$ equations. Computing the derivative with respect
to $g$ of the resulting system, and after making some simple
algebraic manipulations, we get for $p=1$:
\begin{equation}\label{der1}
4 g \sum_{n=1}^{2M_{k}}\left[(\chi_n+a_n S_1) P_{n-1}^{\prime}+
(a_n P_{n-1} )S_1^{\; \prime}\right]=\frac{1}{g},
\end{equation}
and for $1<p\le 2M_k$:
\begin{eqnarray}
S_1^{\; \prime} \left[a_p-2g(M_k+1-p)a_{p-1}-2g
\sum_{i=1}^{p-2}(\chi_{p-i-1}\;\chi_i)\right]+\nonumber \\
 4 g \sum_{n=0}^{2M_k-p}\left[(\chi_{n+p}+a_{n+p}\; S_1)
P_{n}^{\prime}+a_{n+p}\; P_n S_1^{\;
\prime}\right]=\frac{\chi_p+a_p S_1}{g},\label{der2}
\end{eqnarray}
where the primes stand for the derivatives with respect to $g$
and, according to definition (\ref{pn}), we have:
\begin{equation}\label{pnprime}
P_n^{\prime}=\sum_{\beta \notin
C_k}\frac{(n+1)}{(2\eta_k-e_{\beta})^{n+2}}
\left(\frac{de_{\beta}}{dg}\right).
\end{equation}

Taking the limit $g \rightarrow g_c$ and reordering the equations
in a convenient way we  obtain
\begin{equation}\label{der1gc}
\left(-\frac{1}{g_c}+4 g_c \sum_{n=1}^{M_k}\chi_n P_{n-1}^{\prime}
\right)+\sum_{n=2}^{2M_k} (4 P_{n-1}) (a_n S_{1}^{\; \prime})=0 ,
\end{equation}
for $p=1$, and
\begin{eqnarray}
\left(-\frac{\chi_p}{g_c}-2g_c S_1^{\;\prime}
\sum_{i=1}^{p-2}(\chi_{p-i-1}\;\chi_i) +4 g_c
\sum_{n=0}^{M_k-p}\chi_{n+p} P_{n}^{\prime} \right)
-2g_c(M_k+1-p)\;a_{p-1} S_1^{\;\prime} +\nonumber \\ (1+4g_c
P_0)\; a_p S_1^{\;\prime}+
 \sum_{j=p+1}^{2M_k} (4 g_c P_{j-p}) (a_{j} S_1^{\;\prime})=0 ,
\label{der2gc}
\end{eqnarray}
for $1<p\le 2M_k$. This is a non-homogeneous system of $2 M_k$
equations  with a set of $2M_k+(M-M_k)$ unknowns:
$S_1^{\;\prime}$, $a_2$, $a_3$, $\dots$, $a_{2M_k}$ joined with the
derivatives $(de_{\beta}/dg)$, for $\beta \notin C_k$.

The system of equations (\ref{der2gc}) is non-linear  because of
the products $a_{j} S_1^{\;\prime}$. However, we can obtain a linear
system for the derivatives with a unique solution by defining a
matrix $\mathbf{B}$ as:

\begin{align} \nonumber
&B_{p,1}  =\left(-\frac{\chi_p}{g_c}-2g_c S_1^{\;\prime}
\sum_{i=1}^{p-2}(\chi_{p-i-1}\;\chi_i) +4 g_c
\sum_{n=0}^{M_k-p}\chi_{n+p} P_{n}^{\prime} \right) ,\\[3mm]
 &B_{p,p-1}  =-2 g_c (M_k+1-p) , \qquad \qquad B_{p,p} = (1+
4g_c P_0) \quad (p>1) ,
\nonumber  \\[3mm]
& B_{p,j} = 4g_c P_{j-p} \quad (j>p) , \qquad \qquad \quad B_{p,j}
= 0 \quad (1<j<p-1); \nonumber
\end{align}

and the non-null vector $\mathbf{v}$ :
\[
\mathbf{v}=\left(1,\; a_2\;\frac{d\!S_1}{dg},
\;a_3\;\frac{d\!S_1}{dg},\;\dots\;, \;a_{2M_k}\;\frac{d\!S_1}{dg}
\right).
\]

Using these definitions, equations (\ref{der2gc}) can be rewritten
as:
\[ \mathbf{B}\cdot \mathbf{v}=\mathbf{0} , \]
such that  the following condition must be satisfied:
\begin{equation}\label{deter}
\det(\mathbf{B})=0.
\end{equation}

Since the derivatives appear just in the first column of the
matrix $\mathbf{B}$, (\ref{deter}) is a linear equation in
$\frac{d\!S_1}{dg}$ and in the  ($M-M_k$) derivatives
$\frac{de_{\beta}}{dg}$ (for $\beta \notin C_k$) that does not
include the unknown coefficients $a_p$. A set of $(M-M_k)$
equations is required to complete the system. In order to get the
necessary equations we will compute the derivatives of the
Richardson equations out of the cluster:

\begin{equation}
1-4g\sum\limits_{j = 1}^{L}\frac{d_{j}}{2\eta _{j}-e_{\alpha }}+
4g\sum\limits_{\scriptstyle \beta \notin C_k  \atop
  \scriptstyle (\beta \ne \alpha)}\frac{1}{e_{\alpha
}-e_{\beta }} +4g\sum\limits_{\beta  \in C_k}\frac{1}{e_{\alpha
}-e_{\beta }}=0  \qquad  (\alpha \notin C_k) .
\end{equation}
Expanding the last term in the variables $S_p$ we obtain
\begin{equation} \label{noncluster}
1-4g\sum\limits_{j = 1}^{L}\frac{d_{j}}{2\eta _{j}-e_{\alpha }}+
4g\sum\limits_{\scriptstyle \beta \notin C_k  \atop
  \scriptstyle (\beta \ne \alpha)}\frac{1}{e_{\alpha
}-e_{\beta }} -4g\sum_{n=0}^{\infty}\frac{1}{(2\eta_k- e_{\alpha
})^{n+1}}S_n=0 .
\end{equation}

Computing the derivative with respect to $g$ and taking the limit
$g\rightarrow g_c$ we finally get:

\begin{eqnarray}
-\frac{1}{g_c}-4g_c\sum\limits_{j = 1}^{L}\frac{d_{j}}{(2\eta
_{j}-e_{\alpha })^2}\left(\frac{de_{\alpha}}{dg}\right)_{g=g_c}+
4g_c\sum\limits_{\scriptstyle \beta \notin C_k \atop
  \scriptstyle (\beta \ne \alpha)}\frac{1}{(e_{\alpha
}-e_{\beta })^2}\left( \frac{de_{\beta}}{dg}-
\frac{de_{\alpha}}{dg} \right)_{g=g_c} \nonumber \\
-4g_c\frac{M_k}{(2\eta_k- e_{\alpha })^{2}}
\left(\frac{de_{\alpha}}{dg}\right)_{g=g_c}
-4g_c\left(\sum_{n=1}^{M_k}\frac{\chi_n}{(2\eta_k- e_{\alpha
})^{n+1}}
 \right)\frac{d\!S_1}{dg}=0 \qquad (\alpha \notin C_k).
{\label{eders}}
\end{eqnarray}

Equations (\ref{deter}) and (\ref{eders}) constitute a
non-homogenous linear system with a unique solution for the
desire derivatives.

\section{A numerical example: fermions in a
2-dimensional lattice}

In this section we will  apply our formalism to a pairing model of
fermions in a 2-D square lattice of $N \times N$ sites with
periodic boundary conditions. This example was previously treated
in Ref. \cite{duke1} for the ground state with a repulsive pairing
interaction ($g>0$). We will now analyze the model using the
methodology developed here, focussing on the critical $g$ values
and in the behavior of the solutions in their vicinities.

\begin{figure}
\vspace{-0.3cm} \hspace{-.8cm}
\includegraphics[width=9.3cm]{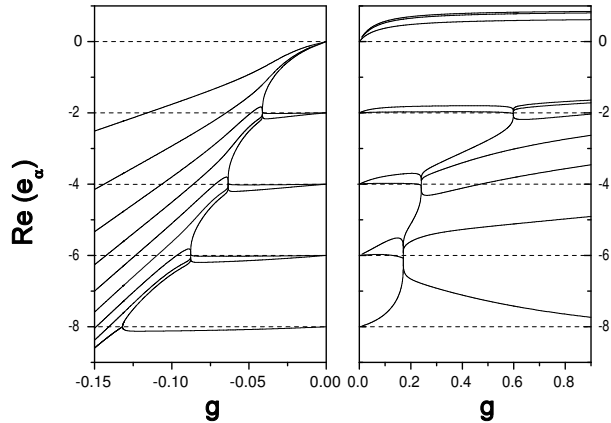}
\caption{Real part of the pair energies $e_\alpha$ for attractive
($g < 0$) and repulsive ($g > 0$) pairing in a $6 \times 6$
lattice.} \label{T1}
\end{figure}

We will consider a $6\times 6$ lattice with $36$ fermions (half
filling, $M=18$). The pairing Hamiltonian in momentum space is
\[
H_P  = \sum\limits_{\textbf{k}} {\varepsilon _\textbf{k} {\kern
1pt} a_{\textbf{k}}^\dagger {\kern 1pt} } a_{\textbf{k}}  +
\frac{g}{2}\sum\limits_{\scriptstyle \textbf{k} \hfill \atop
\scriptstyle \textbf{k}' \hfill} {\kern 1pt}
a_{\textbf{k}}^\dagger {\kern 1pt}
  a_{\overline{\textbf{k}}}^\dagger  {\kern 2pt} a_{\overline{\textbf{k}}'} {\kern 1pt} a_{\textbf{k}'}
\]
where $\textbf{k}\equiv{[k_x,k_y]}$ and $\varepsilon
_\textbf{k}=-2[\cos(k_x)+\cos(k_y)]=\eta_{\textbf{k}}$, and
$\overline{\textbf{k}}$ is the time reversal of $\textbf{k}$. The
single fermion energies $\varepsilon _{j}$ and the corresponding
degeneracies $\Omega_{j}$ for the $6 \times 6$ lattice are
displayed in the Table \ref{niveles}. In the limit $g=0$ the
ground state is obtained by distributing the $M=18$ pairs in the
lowest possible states. In this numerical example we will work
within the fully paired subspace containing the ground state (
$\nu_j=0$, $d_j=-\Omega_j/4$ ). Consequently,
 the excited states to which we will refer to in this example are within the seniority $0$ subspace.

\begin{table}
\caption{Single fermion energies and degeneracies for the $6
\times 6$ lattice.} \label{niveles}
\begin{tabular}
[c]{|c|rrrrrrrrr|}\hline
$j$ & $1$ & $2$ & $3$ & $4$ & $5$ & $6$ & $7$ & $8$ & $9$\\\hline
$\varepsilon_{j}$ & $-4$ & $-3$ & $-2$ & $-1$ & $~0$ & $~1$ & $~2$ & $~3$ &
$~4$\\\hline
$\Omega_{j}$ & $~2$ & $~8$ & $~8$ & $~8$ & $20$ & $~8$ & $~8$ & $~8$ &
$~2$\\\hline
\end{tabular}
\end{table}

We will begin solving the system of equations (\ref{detgc}) for a
specific single particle level, for instance, $j=4$ and
$\varepsilon_4=-1$ (see table \ref{niveles}). The degeneracy of
this level is $8$ and the number of pair energies in the cluster
that collapses at $2\varepsilon_4=-2$ is given by the cluster
condition $M_4=1-2d_4=5$, one more than the value allowed by the
Pauli exclusion principle. The solution of the system
(\ref{detgc}) provides all the information concerning the many
body state at each specific critical value of $g$. In table
\ref{tablaj4} we show, for $j=4$, the first negative $g_c$'s
together with the corresponding energy eigenvalues.

A different kind of analysis is presented in Table \ref{tablags}
where we show the first critical $g$ values for the ground state.
In this case the clusters collapse at different single particle
levels for each  $g_c$ as indicated. All the cluster are formed by
5 pair energies except for $j=1$ where $M_1=3$. In order to have a
global vision of the results we show in Fig. {\ref{T1} the real
part of the pair energies for the ground state solution of the
Richardson equations (\ref{richA}) for positive and negative $g$
values. The full lines are the pair energies $e_\alpha$ and the
horizontal dotted lines correspond to $2\varepsilon_j$.  The
critical points of Table \ref{tablags} can be seen in the figure
at the crossing point of each cluster with twice the corresponding
single particle energy.

The solutions near the critical points $g_c$ were obtained
following the approach described in Section \ref{sectionder}. With
the solution for the first $g_c$ next to $g=0$ (negative or
positive depending on the case), i.e, the $e_{\beta}$ for $\beta
\notin C_k$ and value of $g_c$ itself, we solve the linear system
of equations (\ref{deter}) and (\ref{eders}) to obtain the
derivatives $(dS_1/dg)$ and $(de_{\beta}/dg)$ at the critical
point. In addition we solve the linear system (\ref{linearsys}) to
get the coefficients $\chi_p$. With all this information we
determine a good initial guess at some $g_0$ near $g_c$ by mean of
the linear approximation (\ref{linaprox}). Next we solve the
Richardson equations at $g_0$, a process that rapidly converge to
an accurate numerical solution. With the solution at $g_0$ as a
starting point we move to the next value of $g=g_0 + \Delta g$.
Proceeding in this way and updating the starting values at each
step we compute all the desired points in the curves. We repeat
the process for each critical $g$. The pair energies obtained from
the solutions for $g \approx 0$ smoothly connect with those
obtained near the first $g_c$. The same smooth behavior is
observed between two consecutive critical values.

The good quality of the initial guess near $g_c$ is due to the
smooth behavior of the variables $S_p$ and $e_\beta$  ($\beta
\notin C_k$) that allows us to use the linear approximation
(\ref{linaprox}). For the $e_\beta$'s this fact can be appreciated
in figure \ref{T1}. In figure \ref{T2} we show a graphical
representation of the first $S_p$'s  in a range of $g$ around
$g_c=0.170878$ corresponding to the collapse at the single
particle level $j=2$, \ $2\varepsilon_2=-6$ of a cluster  of
$M_2=5$ pair energies. We can we see in the figure the linearity
of the $S_p$'s for $1\le p \le 5$ confirming that $S_p$ is an
infinitesimal of the same order as $S_1$ for $2 \le p \le M_2$. In
addition, we have plotted the results for $S_6$, which clearly
shows that it is an infinitesimal of a higher order.

\begin{figure}
\vspace{-0.3cm} \hspace{-.8cm}
\includegraphics[width=9.3cm]{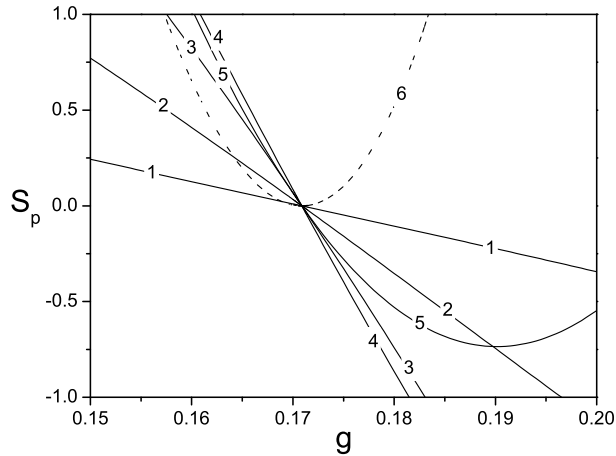}
\caption{Variables $S_p$ close to the collapse at $g_c=0.170878$.
The first 5 $S_p$'s are displayed in solid lines while $S_6$ is
drawn in dashed line. } \label{T2}
\end{figure}

\begin{table}
\caption{First critical $g$ values for the cluster collapsing at
$2\varepsilon_4=-2$} \label{tablaj4}
\begin{tabular}
[c]{|c|l|}\hline
g$_{c}$ & Energy\\\hline
\multicolumn{1}{|l|}{-0.0384565} & -47.6184\\\hline
\multicolumn{1}{|l|}{-0.0391412} & -49.5405\\\hline
\multicolumn{1}{|l|}{-0.0394719} & -53.3549\\\hline
\multicolumn{1}{|l|}{-0.0404240} & -55.5262\\\hline
\multicolumn{1}{|l|}{-0.0412922} & -57.4106\\\hline
\multicolumn{1}{|l|}{-0.0413245} & -62.5795\\\hline
\end{tabular}
\end{table}

\begin{table}
\caption{Critical $g$ values for the ground state} \label{tablags}
\begin{tabular}
[c]{|c|c|c|c|c|}\hline $$ & $2\varepsilon_{4}=-2$ &
$2\varepsilon_{3}=-4$ & $2\varepsilon_{2}=-6$ &
$2\varepsilon_{1}=-8$\\\hline
$g_{c}$ (negative) & -0.0413245 & -0.0635021 & -0.0877434 & -0.131927\\
$g_{c}$ (positive) & 0.598232 & 0.240579 & 0.170878 & ---\\\hline
\end{tabular}
\end{table}
\vspace{5cm}

\section{Conclusions}
In this work we have studied the solution of the Richardson
equations close to the critical values of the coupling constant
$g$. We have derived a set of well behaved equations to determine
the actual critical values $g_c$ associated with any single
particle energy $\varepsilon_k$, and the complete set of pair
energies defining the exact eigenstate of the system. In addition,
we studied the behavior of the solutions close to the critical
points, obtaining a linear approximation for the pair energies
that serves as a good initial guess in the critical region. With
this formalism, one can solve numerically the Richardson equations
around the critical points, overcoming the numerical instabilities
that usually arise. The knowledge of the critical values of the
coupling constant greatly simplifies the numerical treatment of
the equations. To illustrate the formalism we have analyzed the
pairing Hamiltonian in a $6 \times 6$ square lattice at half
filling for repulsive and attractive pairing strength. Making use
of our approach we have located the $g_c$'s for the ground state
and several excited states in the Seniority 0 subspace. The
numerical solution between consecutive values of $g_c$'s can be
easily carried out by solving the original Richardson equations
(\ref{richA}). With this new approach we hope to be able to solve
exactly large systems with arbitrary single particle energies and
degeneracies.

\section*{acknowledgments}
We acknowledge fruitful discussions S. Rombouts. This work was
supported by Spanish DGI under grant BFM2003-05316-C02-02.

\end{document}